\documentclass[letterpaper, 10 pt, conference]{ieeeconf}

\IEEEoverridecommandlockouts                              
\usepackage{amsmath} 
\usepackage{amssymb}  
\usepackage{mathtools}
\usepackage{cite}

\usepackage{color}
\usepackage[colorlinks=true,linkcolor=blue]{hyperref}

\def\E#1{{\mathbb{E}\left[#1\right]}}
\def\Var#1{\textrm{Var}{\left[#1\right]}}

\newtheorem{theorem}{Theorem}

\newtheorem{corollary}{Corollary}

\def\equationautorefname~#1\null{(#1)\null}

\def\qed{ \rule{.08in}{.08in}}

\newenvironment{proof-of}[1]
  {\noindent\textbf{Proof of #1: }}
  {\hspace*{\fill}~\qed\par\endtrivlist\unskip}

\title{\Large \bf Analysis of Difficulty Control in Bitcoin and Proof-of-Work Blockchains\thanks{
This work was supported by National Science Foundation grant no. 16907101.00 and US Air Force grant no. FA9550-16-1-0290.
Daniel Fullmer and A. Stephen Morse are with the Department of Electrical Engineering, Yale University,~\texttt{\{daniel.fullmer,as.morse\}@yale.edu}.
}}

\author{Daniel Fullmer
  \hspace*{0.5 in}
  A. Stephen Morse
}

\begin{document}

\maketitle
\thispagestyle{empty}
\pagestyle{empty}

\begin{abstract}
    This paper presents a stochastic model for block arrival times based on the difficulty retargeting rule used in Bitcoin, as well as other proof-of-work blockchains.
    Unlike some previous work, this paper explicitly models the difficulty target as a random variable which is a function of the previous block arrival times and affecting the block times in the next retargeting period.
    An explicit marginal distribution is derived for the time between successive blocks (the blocktime), while allowing for randomly changing difficulty.
    This paper also aims to serve as an introduction to Bitcoin and proof-of-work blockchains for the controls community, focusing on the difficulty retargeting procedure used in Bitcoin.
\end{abstract}

\section{Introduction}
\textit{Bitcoin} is a decentralized digital currency (or \textit{cryptocurrency}) operated by an ad-hoc network of computers.
It enables peer-to-peer payments without requiring a trusted third party.
Bitcoin's original ``whitepaper''~\cite{Nakamoto2008} was released late 2008, and the currency was launched in 2009.
There has been a significant amount of current interest in Bitcoin and alternative cryptocurrencies, as well as the technology underlying Bitcoin, the ``blockchain.''

This paper focuses on one aspect of Bitcoin and blockchain-based systems, specifically \textit{difficulty retargeting}, also called \textit{difficulty readjustment} or \textit{difficulty control}.
Some existing published analysis of difficulty control in Bitcoin may be found in~\cite{Kraft2016} and~\cite{Bowden2018}.
In~\cite{Bowden2018}, the authors note that the block arrival times do not follow a Poisson distribution
and present a variety of modeling alternatives, testing them against real data from the Bitcoin blockchain.
In~\cite{Kraft2016} the Bitcoin mining process is treated as a nonhomogeneous Poisson process with a deterministic intensity function $\lambda(t)$.
Their analysis principally focuses on the design of a difficulty retargeting algorithm under the assumption of an exponentially increasing hashrate.
However, as noted in the paper,~\cite{Kraft2016}~does not account for the fact that $\lambda(t)$ is itself a stochastic process depending on the time of the arrivals of the process.\footnote{%
Poisson processes whose intensity functions are themselves stochastic processes are sometimes called ``Cox processes''~\cite{Cox1955} or ``doubly stochastic Poisson processes.''
The authors of this paper are not aware of the study of general Poisson processes whose intensity function depends on previous arrivals in the same way as considered in this paper.
}
This paper explicitly considers this case, and furthermore derives a marginal distribution for the time between successive blocks as well as their expected value and variance.
In order to derive these results, a stochastic model for block arrival times is developed as well.

While this paper specifically focuses on the difficulty retargeting rule used in Bitcoin,
the analysis applies to a number of related cryptocurrencies relying on proof-of-work which use a similar retargeting rule.

In \autoref{background}, Bitcoin and blockchains are described and motivation is given for the difficulty adjustment mechanism.
In \autoref{problem-formulation}, the stochastic model for block arrival times is developed.
In \autoref{analysis}, this model is analyzed and the main result of a marginal distribution, expected value, and variance for block times is presented.
Finally, in \autoref{simulations}, simulations of the block arrival process are presented and compared with the analytical results given in the previous section.

\section{Background}\label{background}
The following description of Bitcoin and blockchains omits certain details which are not relevant to the specific problem considered in this paper.
With that said,
it is intended to describe and motivate the purpose of blockchains for cryptocurrencies,
the relevant property proof-of-work blockchains ensure (immutability),
and the purpose of difficulty retargeting.
Readers already familiar these concepts may skip to \autoref{problem-formulation}.
Readers desiring additional details are encouraged to read~\cite{Nakamoto2008,Tschorsch2016a}.

Bitcoin is a cryptocurrency which relies on a public ledger of transactions.
All transactions are recorded on this public ledger, called the ``blockchain''.
This ledger may be thought of as an ordered list of transactions.
Each transaction includes the address of the sender, the recipient, the amount, and a digital signature from the sender.
Senders of the currency can create, sign, and submit transactions to be included on the public ledger.
Recipients can check that transactions are valid and included on the ledger (confirmed).
A transaction is valid if it meets a number of criteria, including, if digital signature is valid and the payer has enough currency (as determined by the history of valid transactions on the ledger before that transaction.)
There also is a special type of transaction for creating new currency in the system.

Previous digital currency systems required a trusted third party intermediary to maintain and publish the ledger.
However, a third party maintainer of the public ledger must always be online and available.
It is a centralized, single point of failure.
Although such a potential maintainer is a trusted third party, it is not able to forge the digital signatures required for valid transactions.
As a result, it cannot arbitrarily transfer funds from one user to another.
It can, however, add, remove, or reorder previous transactions on the ledger, as well as censor transactions from particular users.
The ability to add, remove, or reorder transactions in the past may invalidate later transactions.
As a simple example, if one user transfers some amount of currency to another user to another exchange for some good,
and the maintainer later removes this transaction,
the original user then has both the good as well as his or her original currency,
and the other user has nothing.
So, recipients in the system would want to ensure that previous transactions are unchangeable (\textit{immutable}).
The fundamental innovation of Bitcoin was to create a distributed public ledger which could ensure immutability for past transactions.
This distributed public ledger is the \textit{blockchain}.

In a blockchain, transactions are grouped into blocks.
Each block contains a reference to the previous block, creating a chain.
Every user of the system keeps a copy of the blockchain\footnote{This is not strictly true, but is one of the details which is not relevant for our discussion.}.
New blocks are created by users who decide to participate in \textit{mining}.
These users are called \textit{miners}.
They collect new transactions, attempt to create (\textit{mine}) a new block including those transactions, and publish the newly created block to all other users.
However, mining a new block is intentionally difficult.
It requires a \textit{proof-of-work}~\cite{Dwork,Back2002}.
This proof-of-work can be thought of as a solution to a difficult mathematical puzzle which depends on the data in the candidate block.
The purpose of the proof-of-work is to enable immutability of the blockchain, as described below.

For our purposes, a proof-of-work function is a function $W : \mathcal B \times \mathbb Z \to [0,1]$,
where $B$ is the set of valid blocks, and $\mathbb Z$ is the set of integers.
The goal is to find, given a block $b \in \mathcal B$,  an integer ``nonce'' $\eta$ satisfying the following:
\begin{equation}\label{pow}
    W(b, \eta) < \frac{1}{d}
\end{equation}
for some (large) difficulty target $d > 0$.
The function $W$ is assumed to be a random oracle, which means that each unique evaluation of $W$ produces a random number uniformly in $[0,1]$.
Each miner repeatedly evaluates $W$ with different nonces until they find one satisfying the difficulty target.
Attempts to find a solution are successful with some small probability as determined by $d$.
If they find one, their block with the included nonce is considered valid and will be accepted by others in the network,
and we say that the miner has successfully found a block, or that the miner has \textit{mined} a block.
A block and nonce with a low $W(\cdot)$ value is proof that the miner has done a significant amount of work.
For Bitcoin, this proof-of-work function is based on the SHA-256 hash function~\cite{NIST2002}, but the details of this are not necessary for this paper, and the model in \autoref{pow} will suffice.

Successfully mining a block includes a block reward, which is a special transaction creating a predetermined amount of new currency which is allocated to the miner who successfully mined the block.
As a result, there is significant incentive for each miner to dedicate computational resources to the task of mining.

Since various miners may produce multiple blocks based on the same previous block,
multiple versions of the blockchain may exist simultaneously on the network,
but there is incentive for miners and users to come to a consensus on one version.
The rule which leads to consensus is this: the longest\footnote{More precisely, the canonical blockchain is the one with the most accumulated proof-of-work.} valid blockchain is the canonical one.
A block is valid if all transactions in the block are valid, the previous block is valid, and the proof-of-work is satisfied.
Because of this rule, miners are incentivized to mine new blocks which based on the existing longest blockchain, so that their rewards are accepted by all other users.

As previously mentioned, the proof-of-work system contributes to the immutability of the blockchain.
If, for instance, an adversarial miner wants to remove some transaction in the past,
he or she could create a new version of the blockchain based on the block immediately preceding the targeted transactions with that specific transaction omitted.
However, in order for this new blockchain to be accepted as canonical, it would have to become longer than the existing blockchain.
If the adversary controls less than half of the mining processing power, it will mine blocks less frequently (on average) than the miners mining on top of the existing blockchain.
It is unlikely the adversary's blockchain could surpass the existing blockchain, and for this reason, blocks far in the past are treated as immutable by the users in the system.
For more details, see~\cite{Nakamoto2008}.

Recalling \autoref{pow}, if the difficulty target $d$ is too high or too low, solutions will be found by miners too frequently or infrequently.
The desired goal is to have new solutions (and therefore new blocks) found every $\beta = 10$ minutes, on average.
This parameter $\beta$ was chosen as a tradeoff between ensuring blocks have sufficient time to propagate to all users in the network,
and ensuring that transactions do not take too long to be confirmed (included on the blockchain).
Since miners continue to dedicate additional computational resources to the task of mining, absent any accommodating factor, blocks would be mined too frequently.
So, there is a difficulty retargeting algorithm as part of the Bitcoin consensus rules
which adjusts the difficulty upward if blocks are found too frequently,
or adjusts the difficulty downward if blocks are found too infrequently.
This may be thought of as a \textit{difficulty control problem} integral to blockchains which rely on proof-of-work.

\subsection{Notation}
For a random variable $X$, $X \sim \textrm{Dist}(\cdot)$ is denotes that $X$ is distributed according to some distribution $\textrm{Dist}(\cdot)$.
For a continuous random variable $X$, $f_X(x; \theta)$ represent the probability density function of $X$ parameterized by $\theta$.
The families of probability distributions used in this paper are $\textrm{Exp}(\lambda)$, $\textrm{Erlang}(N, \lambda)$, and $\textrm{Lomax}(N, \lambda)$
which are the exponential, Erlang, and Lomax distributions with rate parameter $\lambda$ and shape parameter $N$.

\section{Problem Formulation}\label{problem-formulation}
Suppose blocks are found at the times given by the random variables $0 \le t_1 \le t_2 \le \ldots$ with the initial block time $t_0 = 0$.
The time between blocks is denoted by $X_k = t_{k} - t_{k-1}$ for $k \ge 1$, and is called the \textit{blocktime} for block $k$.

Recall that we treat the proof-of-work function $W$ as a random oracle, meaning each unique evaluation samples uniformly a real value between 0 and 1.
As a result, the process of repeatedly evaluating $W(\cdot)$ until a nonce is found which satisfies the difficulty target may be thought of as Bernoulli trials.
As such, the number of evaluations needed until a success is found follows a geometric distribution.
The continuous analogue of a geometric distribution is the \textit{exponential distribution}
The limiting behaviour of such a geometric distribution as the number of parallel evaluations and difficulty increases to infinity follows an exponential distribution.
See section 2.2.5 of~\cite{Gallager2013}.


In as similar way as in~\cite{Kraft2016}, for each $k \ge 1$, the random variable $X_k$ is assumed to be distributed according to an exponential distribution with a rate $\lambda_k$ given by
\begin{equation}\label{lambda-defn}
    \lambda_k = \frac{r_k}{d_k}
\end{equation}
where $d_k$ and $r_k$ are two positive real (random) variables, called the \textit{difficulty} and the \textit{hashrate} respectively.
The hashrate may be thought of as representing the sum of the computational resources dedicated toward mining at that time.
This is determined exogenously by the miners.
The difficulty, however, is updated automatically according to the Bitcoin consensus rules.
Recall that the expected value of an exponentially distributed random variable is equal to the inverse of its rate.
So, given a known $\lambda_k$, $\E{X_k | \lambda_k} = 1/\lambda_k$.

The design of Bitcoin includes a ``difficulty retargeting'' process which periodically updates the difficulty as the hashrate increases or decreases.
The goal is to have a new block found according to a desired blocktime $\beta = 10$ minutes (in expectation).
The difficulty is adjusted according to\footnote{The update rule used in Bitcoin additionally restricts $d_{k+1}$ to only change by a factor of $4$ in either direction. Moreover, the Bitcoin code includes a well-known bug which excludes the final $X_{k}$ in the sum.}
\begin{equation}\label{d-update}
    d_{k+1} = \begin{cases}
        \frac{N \beta}{\sum_{i=1}^N X_{k-N+i}} d_{k} & \textrm{if } k \bmod N = 0 \\
        d_{k} & \textrm{otherwise}
    \end{cases}
\end{equation}
where $N = 2016$ is the number of blocks in each \textit{difficulty retargeting period}
and $d_1$ is assumed to be initialized arbitrarily.
Note that the difficulty is constant between difficulty readjustments.
Intuitively, if the time to mine the previous $N$ blocks took longer than $N \beta$, then the difficulty is decreased.
Likewise, if the time to mine the previous $N$ blocks was shorter than $N \beta$, then the difficulty is increased.

In this paper, for simplicity, we additionally suppose that $r_k$ remains constant during each retargeting period.

We concerned with computing the marginal distribution of the blocktimes $X_k, k \ge 1$,
along with the expected value and variance of block times while accounting for randomly varying difficulty according to \autoref{d-update}.
These results may be found in the sequel as \autoref{thm:main} and \autoref{E-var}.

\subsection{Derivation of adjustment algorithm}
Below is a description of how such a rule \autoref{d-update} might be derived.
Specifically, it's designed so that $\lambda_k^{-1}$ (the expected time to mine the $k$th block) is approximately $\beta$,
assuming the hashrate is unchanging from the previous to the next period.
To derive this update rule, we first attempt to estimate the hashrate in the previous period, $r_k$, knowing only $d_k$ and the previous $X_{k-N+i},\ 1 \le i \le N$.
Toward this end, we estimate the $\lambda_k$, and call it $\hat \lambda_k$, by setting the expected time to mine $N$ blocks equal to the actual time to mine $N$ blocks.
\begin{equation}
    N \E{X_k | \hat \lambda_k} = \frac{N}{\hat \lambda_k}  = \sum_{i=1}^n X_{k-N+i}
\end{equation}
Let $\hat \lambda_k = \hat r_k / d_k$, where $\hat r_k$ is the estimate of the hashrate in the previous period.
\begin{equation}\label{hat-r_k}
    \hat r_k = \hat \lambda_k d_k = \frac{N d_k}{\sum_{i=1}^n X_{k-N+i}}
\end{equation}
With this estimate of the hashrate, the goal is to set $d_{k+1}$ such that the expected blocktime of the next block $X_{k+1}$ is equal to $\beta$, with $r_{k+1}$ assumed to be equal to $\hat r_k$.
\begin{equation}
    \beta = \E{X_{k+1} | \lambda_{k+1}} = \frac{1}{\lambda_{k+1}} = \frac{d_{k+1}}{r_{k+1}} = \frac{d_{k+1}}{\hat r_{k}}
\end{equation}
From this and \autoref{hat-r_k},
\begin{equation}
    d_{k+1} = \frac{N \beta}{\sum_{i=1}^n X_{k-N+i}} d_{k}
\end{equation}
which matches the update rule in \autoref{d-update}.

\section{Analysis}\label{analysis}
Since $d_k$ and $r_k$ are assumed to be constant during each retargeting period,
it proves convenient to introduce the following notation.
\begin{align}
    \bar d_n = d_{(n-1)N + 1} = d_{(n-1)N + 2} = \cdots = d_{nN} \\
    \bar r_n = r_{(n-1)N + 1} = r_{(n-1)N + 2} = \cdots = r_{nN} \\
    \bar \lambda_n = \lambda_{(n-1)N + 1} = \lambda_{(n-1)N + 2} = \cdots = \lambda_{nN} \\
    T_n = \sum_{k=1}^N X_{(n-1)N+k} \label{T_n-defn}
\end{align}
for each $n \ge 1$.
From this and \autoref{d-update} it follows that
\begin{equation}\label{bar-d-update} 
    \bar d_{n+1} = \frac{N \beta}{T_n} \bar d_n, \quad n \ge 1
\end{equation}
From this and $\autoref{lambda-defn}$, for each $n \ge 1$
\begin{equation}\label{bar-lambda-update-deriv} 
    \bar \lambda_{n+1}
    = \frac{\bar r_{n+1}}{\bar d_{n+1}}
    = \frac{\bar r_{n+1} T_n}{N \beta \bar d_n}
    = \frac{\bar r_{n+1} \bar r_n T_n}{\bar r_n N \beta \bar d_n}
    = \delta_{n+1} \frac{T_n}{N \beta} \bar \lambda_n
\end{equation}
where $\delta_n = \bar r_n / \bar r_{n-1}$ for $n > 1$.
It proves convenient to define $\theta_n = \frac{N \beta}{\delta_n}$ for $n \ge 1$.
So,
\begin{equation}\label{bar-lambda-update}
    \bar \lambda_{n+1} = \frac{T_n}{\theta_{n+1}} \bar \lambda_n
\end{equation}
for each $n \ge 1$.

So for each $n \ge 1$ and $1 \le k \le N$, the block time $X_{(n-1)N + k}$ is exponentially distributed according to $\bar \lambda_n$.
However, while $\bar \lambda_1$ is a fixed value, each $\bar \lambda_n,\ n > 1$ is a random variable.
In other words:
\begin{align}
    X_k &\sim \textrm{Exp}(\bar \lambda_1) & 1 \le k \le N \\
    X_{(n-1)N + k} | \bar \lambda_n &\sim \textrm{Exp}(\bar \lambda_n) & n > 1,\ 1 \le k \le N 
\end{align}
So, the (conditional) probability density functions are as follows:
\begin{align}
    f_{X_k}(x) &= \bar \lambda_1 e^{-\bar \lambda_1 x} & 1 \le k \le N \label{X_k-pdf} \\
    f_{X_{(n-1)N + k} | \bar \lambda_n}(x, \lambda) &= \lambda e^{-\lambda x} & n > 1,\ 1 \le k \le N \label{X_k-cpdf}
\end{align}
Here the distribution of each $X_{(n-1)N + k},\ n > 1,\ 1 \le k \le N$ is conditional on the value of $\bar \lambda_n$.
Since each $T_n$ is the sum of $N$ i.i.d, exponentially distributed random variables whose parameter is $\bar \lambda_n$,
$T_n$ follows an Erlang distribution with parameters $N$ and $\bar \lambda_n$.
Similarly,
\begin{align}
    T_1 &\sim \textrm{Erlang}(N, \bar \lambda_1) \\
    T_n | \bar \lambda_n &\sim \textrm{Erlang}(N, \bar \lambda_n) & n > 1 \\
    f_{T_1}(t) &= \frac{\bar \lambda_1^N t^{N-1} e^{-\bar \lambda_1 t}}{(N-1)!} \label{T_1-pdf} \\
    f_{T_n | \bar \lambda_n}(t, \lambda) &= \frac{\lambda^N t^{N-1} e^{-\lambda t}}{(N-1)!} & n > 1 \label{T_n-cpdf}
\end{align}

We next derive the conditional density function for $\bar \lambda_{n+1} | \bar \lambda_n,\ n \ge 1$.
Since $\bar \lambda_{n+1}$ is monotonically increasing in $T_n$, we can perform a change of variables from $\bar \lambda_{n+1}$ to $T_n$ 
to determine the p.d.f.~of $\bar \lambda_{n+1}$ conditioned on $\bar \lambda_n$.
From \autoref{bar-lambda-update},
\begin{align*}
    f_{\bar \lambda_{n+1} | \bar \lambda_n}(\lambda' | \lambda)
    &= \left( \frac{d}{d \lambda'} \frac{\theta_{n+1} \lambda'}{\lambda} \right) f_{T_n | \bar \lambda_n} \left(\frac{\theta_{n+1} \lambda'}{\lambda}, \lambda \right)
\end{align*}
From this and \autoref{T_n-cpdf},
\begin{align*}
    f_{\bar \lambda_{n+1} | \bar \lambda_n}(\lambda' | \lambda)
    &= \frac{\theta_{n+1}}{\lambda} \frac{\lambda^N \left( \theta_{n+1} \frac{\lambda'}{\lambda} \right)^{N-1} e^{-\theta_{n+1} \lambda'} }{(N-1)!} \\
    &= \frac{\theta_{n+1}^N {(\lambda')}^{N-1} e^{-\theta_{n+1} \lambda'} }{(N-1)!}
\end{align*}
Note two things:
First, this is an Erlang distribution with parameters $N$ and $\theta_n$.
Second, this expression is independent of $\lambda$.
Writing this more succinctly,
\begin{equation}\label{bar-lambda-dist}
    \bar \lambda_n \sim \textrm{Erlang}(N, \theta_n)  \quad \textrm{where} \quad \theta_n = \frac{N \beta}{\delta_n}, \quad n > 1.
\end{equation}
Knowing this, the expected value can be easily calculated as
\begin{equation}
    \E{\bar \lambda_n} = \frac{N}{\theta_n} = \frac{\delta_n}{\beta}
\end{equation}
for $n > 1$.

Using this, it is possible to derive the p.d.f.\ for all $X_k$ beyond the initial period.
These random variables have distributions whose parameters are themselves random variables,
which are referred to as \textit{compound distributions}~\cite{Johnson1994}.
Specifically, the distribution for $X_{(n-1)N + k},\ n > 1,\ 1 \le k \le N$ is a \textit{Lomax distribution},
which is the result of compounding an exponential distribution \autoref{X_k-cpdf} with its rate parameter $\bar \lambda_n$ set according to an Erlang distribution \autoref{bar-lambda-dist}.
Computing the p.d.f.\ for $X_{(n-1)N + k},\ n > 1,\ 1 \le k \le N$ using \autoref{X_k-cpdf} and \autoref{bar-lambda-dist} gives:
\begin{align*}
    f_{X_{(n-1)N+k}}(x)
    = \int_{\lambda=0}^\infty f_{X_{(n-1)N+k} | \bar \lambda_n}(x, \lambda) f_{\bar \lambda_n}(\lambda) d\lambda
\end{align*} 
\begin{align*}
    \quad
    &= \int_{\lambda=0}^\infty \lambda e^{-\lambda x} \frac{\theta_n^N \lambda^{N-1} e^{-\lambda \theta_n}}{(N-1)!} d\lambda \\
    &= \frac{N \theta_n^N}{(x+\theta_n)^{N+1}} \int_{\lambda=0}^\infty \frac{(x + \theta_n)^{N+1} \lambda^N e^{-(x + \theta_n) \lambda}}{N!} d\lambda \\
    &= \frac{N \theta_n^N}{(x+\theta_n)^{N+1}} \int_{\lambda=0}^\infty \textrm{Erlang}(\lambda; N+1, x + \theta_n) d\lambda \\
    &= \frac{N \theta_n^N}{(x+\theta_n)^{N+1}}
\end{align*}
This gives our main result:
\begin{theorem}\label{thm:main}
Using the difficulty retargeting rule in \autoref{d-update},
for $n > 1,\ 1 \le k \le N$ the marginal distribution of $X_{(n-1)N + k}$ is the Lomax distribution with parameters $N$ and $\theta_n$.
i.e.
\begin{equation*}
    X_{(n-1)N + k} \sim \textrm{Lomax}(N, \theta_n), \quad n > 1,\ 1 \le k \le N
\end{equation*}
\vspace*{0.1cm}
\end{theorem}

The expected blocktime and variance are easily computed knowing this distribution.
\begin{corollary}\label{E-var}
Using the difficulty retargeting rule in \autoref{d-update},
for $n > 1,\ 1 \le k \le N$
\begin{equation*}
    \E{X_{(n-1)N + k}} = \frac{\theta_n}{N-1} = \frac{N}{(N-1) \delta_n} \beta
\end{equation*}
assuming $N > 1$.
Additionally,
\begin{align*}
    \Var{X_{(n-1)N + k}}
    &= \frac{\theta_n^2 N}{(N-1)^2 (N-2)} \\
    &= \frac{N^3}{(N-1)^2(N-2) \delta_n^2} \beta^2
\end{align*}
assuming $N > 2$.
\end{corollary}

If, instead, the difficulty $d_k$ was assumed to be constant, each blocktime would indeed be distributed according to an exponential distribution with fixed rate parameter $\lambda = 1/\beta$,
whose expected value would be $\beta$ and variance would be $\beta^2$.
It is clear that the difficulty retargeting procedure in \autoref{d-update} leads to slightly higher expected value and variance.
So, even in the case of constant hashrate $\delta_n = 1$, the Bitcoin blockchain runs too fast by a factor of $N/(N-1)$.
However, for the value of $N$ used in Bitcoin, 2016, $N/(N-1)$ is very close to $1$.

One modification to \autoref{d-update} which would provide slightly better results would be to change $N$ to $(N-1)$.
With this modification, $\theta_n = (N-1) \beta / \delta_n$.
And, supposing $\delta_n = 1$, the expected value of $X_{(n-1)N + k},\ n > 1, 1 \le k \le N$ is just $\beta$, as desired, and its variance is $\frac{N}{N-2} \beta^2$.

\section{Simulations}\label{simulations}
\begin{figure}[!t]
    \centering
    \includegraphics[width=0.5\textwidth]{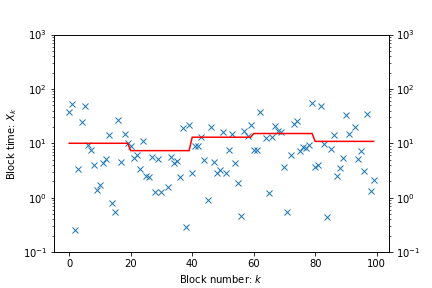}
    \caption{Sampling with $N=20$}\label{fig:N20}
\end{figure}

\begin{figure}[!t]
    \centering
    \includegraphics[width=0.5 \textwidth]{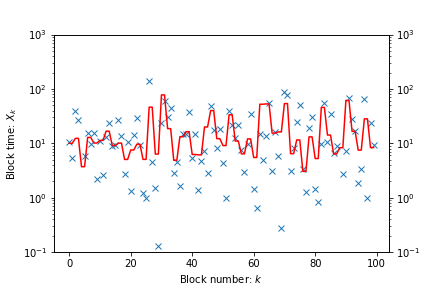}
    \caption{Sampling with $N=2$}\label{fig:N2}
\end{figure}

In this section, we sample a realization of the random variables $X_k$ and $d_k$ for $k \ge 1$ and different values of the parameter $N$.
\autoref{fig:N20} and \autoref{fig:N2} are two realizations of the stochastic process $X_k$, for $N=2$ and $N=20$ respectively.
The blue `x's represent individual block times $X_k$, and the red line represents the value of $1/\lambda_k$, which is the expected value of $X_k$.
Note that the y-axes used in these figures are logarithmic.

In these simulations the parameter $\beta$ is set to $10$ and $\lambda_1$ set to $1/10$, so the blocktimes in the initial period have an expected value of $10$.
For \autoref{fig:N2}, $\lambda_k$ is adjusted every other unit of time based on the values of $X_k$ for the previous two blocks, which leads to significantly more variation in the value of $\lambda_k$, as compared with \autoref{fig:N20}.
The quality of the difficulty adjustment algorithm may be intuitively evaluated by how closely the red line stays to the value $10$.
As can be seen by \autoref{E-var}, the variance of these block times becomes particularly bad for small values of $N$.
In fact, for $N=2$, the blocktimes have infinite variance, as a result of them being Lomax-distributed.

\section{Conclusion}
Future work may consider additional difficulty retargeting rules used in other cryptocurrencies,
as well as studying the interaction between multiple blockchains which share a common proof-of-work scheme.


\addtolength{\textheight}{-12cm}   

\end{document}